\documentclass[aps,twocolumn,superscriptaddress,floatfix,preprintnumbers]{revtex4-1}
\usepackage[utf8]{inputenc}  
\usepackage{amsmath}
\usepackage{amssymb} 
\usepackage{multirow}

\usepackage{enumerate}
\usepackage{bigints}  
\usepackage{tikz}
\usepackage{tipa}
\usepackage{CJKutf8}
\usepackage{feynmf}
\usepackage{slashed}
\usepackage{braket}
\usepackage[toc,page]{appendix}
\usepackage{url}
\usepackage{natbib}
\usepackage{graphicx}
\usepackage[pdftex,bookmarks,linktocpage,pdfpagelabels,plainpages=false,hyperfigures,linkcolor=blue,citecolor=blue]{hyperref} 
\hypersetup{colorlinks=true}

\usepackage{mathrsfs,amssymb}  
\usepackage{cancel}
\usepackage[normalem]{ulem}
\usepackage{array}
\usepackage{booktabs}
\usepackage{verbatim}
\usepackage{kantlipsum}

\begin{document}

\author{Daeyeong Jeong}
\email{jeong.physics@gmail.com}
\affiliation{Department of Physics and Institute for Sciences of the Universe, Chungnam National University, Daejeon 34134, Republic of Korea}

\author{Doojin Kim}
\email{doojin.kim@usd.edu}
\affiliation{Department of Physics, University of South Dakota, Vermillion, SD 57069, USA}

\author{Jong-Chul Park}
\email{jcpark@cnu.ac.kr}
\affiliation{Department of Physics and Institute for Sciences of the Universe, Chungnam National University, Daejeon 34134, Republic of Korea}

\title{Extracting Dark-Matter Mass from Angular Scanning
}

\begin{abstract} 
We propose a novel method to determine the mass scale of ambient dark matter, applicable to (at least effectively) two-dimensional direct detection experiments that allow for directionality observables. 
Due to the motion of the solar system and Earth relative to the Galactic Center and the Sun, the dark-matter flux exhibits a directional preference. 
We first demonstrate that dark-matter event rates depend non-trivially on the angle between the detection plane and the overall dark-matter flow, with the curvature of this angular spectrum encoding mass information.
As proof of principle, we take the recently proposed Graphene-Josephson-Junction-based superlight dark-matter detector as a concrete example and validate these theoretical expectations through numerical analyses. 
\end{abstract}

\maketitle

\noindent {\bf Introduction.} The dark-matter puzzle is one of the key observational motivations for the new physics beyond the Standard Model (SM). 
While weakly interacting massive particles are well-motivated and popular dark-matter candidates, the experimental efforts in searching for them through their hypothetical non-gravity-based interactions with the SM particles are yet unfruitful. 
Given this situation, alternative dark-matter scenarios are garnering great attention and being (re)visited. 
Among these, the detection of superlight dark matter in the keV-to-MeV mass range has become a rapidly growing field~\cite{Hochberg:2015pha,Hochberg:2015fth,Hochberg:2016ntt,Cavoto:2017otc,Hochberg:2017wce,Baracchini:2018wwj,Hochberg:2019cyy,Blanco:2019lrf,Schutz:2016tid,Knapen:2016cue,Maris:2017xvi,Essig:2019kfe,Knapen:2017ekk,Griffin:2018bjn,Blanco:2021hlm,Blanco:2022cel,Blanco:2022pkt,Romao:2023zqf,Catena:2023qkj,Catena:2023awl}, as advances in device technologies---such as Transition Edge Sensor~\cite{doi:10.1063/1.1770037}, Superconducting Tunnel Junction~\cite{1979ApPhL..34..347D}, Superconducting Nanowire Single-Photon Detector~\cite{doi:10.1063/1.1388868}, Microwave Kinetic Inductance Detector~\cite{Day2003}, and 
Graphene Josephson Junction (GJJ)~\cite{GJJ}---have now reached the sensitivity required to detect the tiny energy deposits left by superlight dark matter.

Many such devices operate on an ``on-off'' type working principle in that if an energy deposit is greater than their detection threshold, a dark-matter event occurrence is recognized, but the magnitude of the energy deposit itself is not or poorly measured.
Since conventional dark-matter direct detection experiments rely on the differential recoil energy spectrum to infer the dark-matter mass scale, this on-off characteristic poses a challenge for superlight dark-matter detection experiments in determining the mass value. 
To address this issue, one possible approach is to vary the detection threshold energy and measure event rates, provided the threshold is adjustable.

In this Letter, we propose an alternative method for determining the dark-matter mass, specifically suited for experiments utilizing an (at least, effectively) two-dimensional detector. 
Since their experimental signatures are closely tied to the behavior of dark-matter scattering targets along the detection plane, the incident angle of a dark-matter particle relative to the plane-normal direction affects the event rate. 
Nevertheless, usual dark-matter velocity distributions assume an isotropic flux, i.e., no directional preference, and therefore, the overall angular dependence would be washed out.
However, since the solar system and the Earth orbit the Galactic Center (GC) and the Sun, respectively, the resulting dark-matter flux exhibits an average directional preference. 
As a consequence, event rates should exhibit a non-trivial dependence on the angle between the detection plane and the overall dark-matter flow, commonly referred to as the dark-matter wind. 
Notably, the velocity component parallel to the detection plane dominates in determining event rates over the component normal to the plane. 
Moreover, dark-matter mass is another key factor in determining event rates when a non-zero threshold energy is present. 
For heavy dark matter, a small velocity is sufficient to exceed the threshold, leaving a detectable signature, while for lighter dark matter, a larger velocity is required.    
Basically, an experiment would {\it actively} rotate the two-dimensional detection plane to expose it to the dark-matter wind at different angles. 
The resultant angular distribution of event rates per unit exposure time then provides a way to infer the dark-matter mass scale. 

We will demonstrate the expectations delineated thus far, taking the GJJ-based superlight dark-matter detector. 
We, however, emphasize that the idea proposed here is broadly applicable to other (effectively) two-dimensional dark-matter detectors. Examples include NEWSdm~\cite{NEWSdm:2017efa}, DIAMOND detectors~\cite{Marshall:2020azl}, and DNA detectors~\cite{Drukier:2012hj, OHare:2021cgj}.
Some of these are capable of measuring the magnitude of the recoil energy.
In such cases, the mass value extracted from the aforementioned angular correlation can be cross-checked with the mass value obtained from the differential recoil energy spectrum. 

\medskip

\noindent {\bf Angular dependence of event rates.}
Given a detector where a dark-matter particle of mass $m_\chi$ is expected to scatter off a scattering target $T$ with a cross-section $\sigma_{\chi T}$, the number of dark-matter events per unit run time per unit detector mass $n_{\rm eve}$ is given by
\begin{equation}
    n_{\rm eve} = \int dE_r dv_\chi f(v_\chi)\frac{d}{dE_r}\left(\langle \overline{N}_T \sigma_{\chi T} v_{\rm rel}\rangle \frac{\rho_\chi}{m_\chi} \right), \label{eq:everate}
\end{equation}
where $\overline{N}_T$ and $\rho_\chi$ are the number of scattering targets in the detector per unit detector mass and the dark-matter energy density near the Earth, respectively. 
$f(v_\chi)$ describes the velocity profile of dark matter for which a modified Maxwell-Boltzmann distribution~\cite{Smith:2006ym} is usually adopted, 
\begin{equation}
    \frac{f(v_\chi)}{4\pi} = \frac{v_\chi^2 \exp\left(-\frac{v_\chi^2}{v_0^2} \right)}{\pi^{3 \over 2}v_0^3\left[{\rm erf}\left(\frac{v_{\rm esc}}{v_0}\right)-\frac{2}{\sqrt{\pi}}\frac{v_{\rm esc}}{v_0}\exp\left(-\frac{v_{\rm esc}^2}{v_0^2} \right) \right]}\,, \label{eq:ModMB}
\end{equation}
with root-mean-square velocity $v_0=220$~km$\cdot$s$^{-1}$ and escape velocity $v_{\rm esc}=550$~km$\cdot$s$^{-1}$.
$v_{\rm rel}$ is the velocity of incident dark matter relative to the scattering target. 

As mentioned earlier, if the detector of interest is {\it two}-dimensional, the velocity component parallel to the detection plane primarily determines the event rate. 
In contrast, the normal component makes subleading or even vanishing contributions, depending on the signal detection principle details. 
For simplicity, we henceforth consider the situation where the normal component is irrelevant unless stated otherwise---this is precisely the case for the GJJ-based detector~\cite{Kim:2020coy}, but one can envision an effective parallel component incorporating small corrections from the normal component. 
The velocity profile in Eq.~\eqref{eq:everate} is now replaced with a parallel component profile, denoted as $\tilde{f}(v_{\chi\parallel})$, and the relative velocity $v_{\rm rel}$ should be evaluated with respect to the parallel component, 
\begin{equation}
    n_{\rm eve}= \frac{\rho_\chi}{m_\chi}\int dE_r dv_{\chi\parallel}\tilde{f}(v_{\chi\parallel})\frac{d\langle \overline{N}_T \sigma_{\chi T} v_{{\rm rel}\parallel}\rangle}{dE_r}\,. 
\end{equation}
One can take the following plane-projection procedure of $f(v_\chi)$ to find $\tilde{f}(v_{\chi\parallel})$~\cite{Kim:2020coy} as long as the contributions from the normal component are negligible:
\begin{equation}
    \tilde{f}(v_{\chi\parallel})=\bigintssss_{-\sqrt{1-\left(v_{\chi\parallel}/v_{\rm esc}\right)^2}}^{{\sqrt{1-\left(v_{\chi\parallel}/v_{\rm esc}\right)^2}}} d \cos \theta \frac{1}{2}\frac{1}{\sin \theta}f\left(\frac{v_{\chi\parallel}}{\sin \theta} \right)\,, \label{eq:ppMBMod}
\end{equation}
where $\theta$ is the angle between the incident dark-matter momentum direction and the detection-plane-normal direction, 
\begin{equation}
    v_\chi \sin\theta = v_{\chi\parallel}\,.
\end{equation}
Since $v_\chi < v_{\rm esc}$, a given $v_{\chi\parallel}$ takes the $\theta$ contributions up to $v_{\chi\parallel}=v_{\rm esc}\sin\theta$ that sets the upper and lower limits of the above integral. 

\begin{figure}[t]
    \centering
    \includegraphics[width=0.7\linewidth]{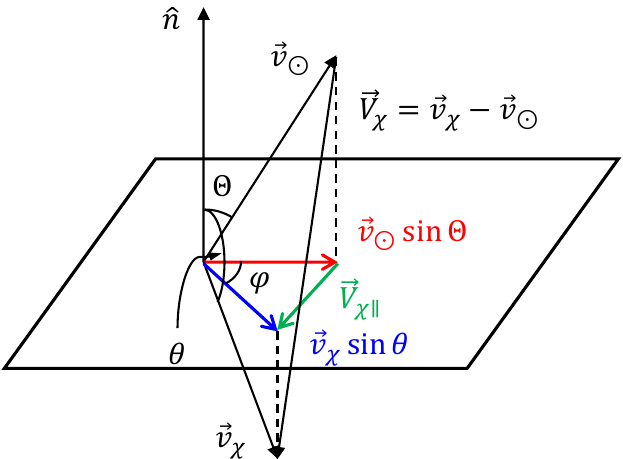}
    \caption{An example configuration of the velocity vectors involved, where $\hat{n}$ is aligned with the normal direction of the detector plane of interest.}
    \label{fig:coord}
\end{figure}

Note that $v_\chi$ is measured with respect to the GC and the formalism thus far is based on the assumption that the dark-matter flux is approximately isotropic. 
Therefore, velocity profiles exhibit no directional dependence. 
However, the solar system orbits the GC, the Earth revolves around the Sun, and the Earth itself rotates. 
As a result, the dark-matter flux or dark-matter wind shows a particular preference toward a certain direction. 
For simplicity, we consider the dominant effect---the revolution of the solar system around the GC---where the dark-matter wind predominantly originates from the direction of the Cygnus constellation. 
Within this framework, we apply the following replacement:
\begin{equation}
    f(v_\chi) \to F(V_\chi)\,,
\end{equation}
where $V_{\chi}\equiv \left| \vec{v}_\chi -\vec{v}_\odot \right|$ is the dark-matter speed relative to the solar system, {\it implicitly} incorporating directional dependence through $\vec{v}_\odot$, which represents the velocity of the solar system toward Cygnus measured with respect to the GC (see also FIG.~\ref{fig:coord}). 
If the detection plane is parallel (normal) to the Cygnus direction, more (less) dark-matter particles are likely to deposit sufficient energy to overcome the detection threshold $E_{\rm th}$. 
In general, for any angle $\Theta$ between the Cygnus direction and the detection-plane-normal direction, a plane-projection procedure of $F(V_\chi)$ should be performed individually and thus $n_{\rm eve}$ depends on $\Theta$ non-trivially:
\begin{equation}
    n_{\rm eve}(\Theta)\propto \int dE_r d V_{\chi \parallel} \tilde{F}(V_{\chi \parallel};\Theta)\frac{d\langle \overline{N}_T \sigma_{\chi T} V_{{\rm rel}\parallel}(\Theta)\rangle}{dE_r}\,,
\end{equation}
where we explicitly show the $\Theta$-dependence of the plane-projected velocity profile $\Tilde{F}$ and the relative velocity $V_{{\rm rel}\parallel}$.

A non-zero $E_{\rm th}$ implies the existence of a minimum $V_{\chi\parallel}$ required for dark-matter signal detection. 
As discussed earlier, if the dark matter of interest is heavy, even a small $V_{\chi\parallel}$ can induce sufficient energy transfer to exceed $E_{\rm th}$. 
In contrast, if $m_\chi$ is small, a relatively large $V_{\chi\parallel}$ is necessary to generate a detectable signal. 
Thus, one can expect a {\it non-trivial dependence of $n_{\rm eve}$ on $m_\chi$ through $V_{\chi\parallel,\min} = V_{\chi\parallel,\min}(m_\chi)$ which influences the curvature of the $\Theta$ dependence}:
\begin{eqnarray}
    n_{\rm eve}(\Theta,m_\chi) &=& \frac{\rho_\chi}{m_\chi}\int dE_r \int_{V_{\chi\parallel,\min}(m_\chi)}^{V_{\chi\parallel,\max}} dV_{\chi\parallel} \tilde{F}(V_{\chi \parallel};\Theta) \nonumber \\
    &\times&  \frac{d\langle \overline{N}_T \sigma_{\chi T} V_{{\rm rel}\parallel}(\Theta)\rangle}{dE_r}\,, \label{eq:master}
\end{eqnarray}
where the maximum velocity is given by $V_{\chi\parallel,\max} = v_{\rm esc} + v_\odot\sin \Theta$. 

Given the configuration in FIG.~\ref{fig:coord}, one can find that $V_{\chi\parallel}$ and $v_\chi$ are related by 
\begin{equation}
    v_\chi\sin\theta = v_{\odot\parallel}\cos\varphi \pm \sqrt{V_{\chi\parallel}^2-v_{\odot\parallel}^2\sin^2\varphi} \equiv \overline{V}_{\chi\parallel}^\pm\,,\label{eq:vchisin}
\end{equation}
where $v_{\odot\parallel}$ is the projection of $\vec{v}_{\odot}$ onto the detector plane:
\begin{equation} 
v_{\odot\parallel}=v_\odot |\sin\Theta|\,,
\end{equation}
which explicitly carries the dependence on $\Theta$.
Here, two solutions arise, which we define as $\overline{V}_{\chi\parallel}^\pm$ for later use. 
It is important to note that $v_\chi\sin\theta$ in Eq.~\eqref{eq:vchisin} must be positive. 
If $V_{\chi\parallel} \geq v_{\odot\parallel}$, only $\overline{V}_{\chi\parallel}^+$ provides a physical solution for any azimuth $\varphi$. 
Conversely, if $V_{\chi\parallel} < v_{\odot\parallel}$, both $\overline{V}_{\chi\parallel}^\pm$ can yield physical solutions for $\pi-\sin^{-1}\left(\dfrac{V_{\chi\parallel}}{v_{\odot\parallel}}\right) \leq |\varphi| \leq \pi$. 
Using the Jacobian factor $J=\left| \dfrac{dv_\chi}{dV_{\chi\parallel}} \right|$ derived from Eq.~\eqref{eq:vchisin}, we finally obtain
\begin{widetext}
\begin{equation}
    \tilde{F}(V_{\chi \parallel};\Theta)=\left\{ 
    \begin{array}{l l}
    \bigintsss_0^\pi d\varphi \dfrac{\tilde{v}_{\parallel}}{\pi} \bigintsss_{-c_\theta^+}^{c_\theta^+}d\cos\theta\dfrac{1}{2\sin\theta}f\left(\dfrac{\overline{V}_{\chi\parallel}^+}{\sin\theta} \right) & \hbox{for }V_{\chi\parallel}\geq v_{\odot\parallel}\cr \cr
    \bigintsss_{\pi-\sin^{-1}\left( \frac{V_{\chi\parallel}}{v_{\odot\parallel}}\right)}^\pi d\varphi \dfrac{\tilde{v}_{\parallel}}{\pi} 
    \left[ \bigintsss_{-c_\theta^+}^{c_\theta^+}d\cos\theta\dfrac{1}{2\sin\theta}f\left(\dfrac{\overline{V}_{\chi\parallel}^+}{\sin\theta} \right) +\bigintsss_{-c_\theta^-}^{c_\theta^-}d\cos\theta\dfrac{1}{2\sin\theta}f\left(\dfrac{\overline{V}_{\chi\parallel}^-}{\sin\theta} \right) \right] & \hbox{for }V_{\chi\parallel}<v_{\odot\parallel}\,,
    \end{array}\right. \label{eq:AngModMB}
\end{equation}
\end{widetext}
where $\tilde{v}_{\parallel}$ and $c_\theta^\pm$ are defined as
\begin{eqnarray}
    \tilde{v}_\parallel &\equiv& \frac{V_{\chi\parallel}}{\sqrt{V_{\chi\parallel}^2-v_{\odot\parallel}^2\sin^2\varphi}}\,, \\
    c_\theta^\pm &\equiv& \sqrt{1-\left(\frac{\overline{V}_{\chi\parallel}^\pm}{v_{\rm esc}}\right)^2}\,.
\end{eqnarray}

\begin{figure}[t]
    \centering
    \includegraphics[width=0.48\textwidth]{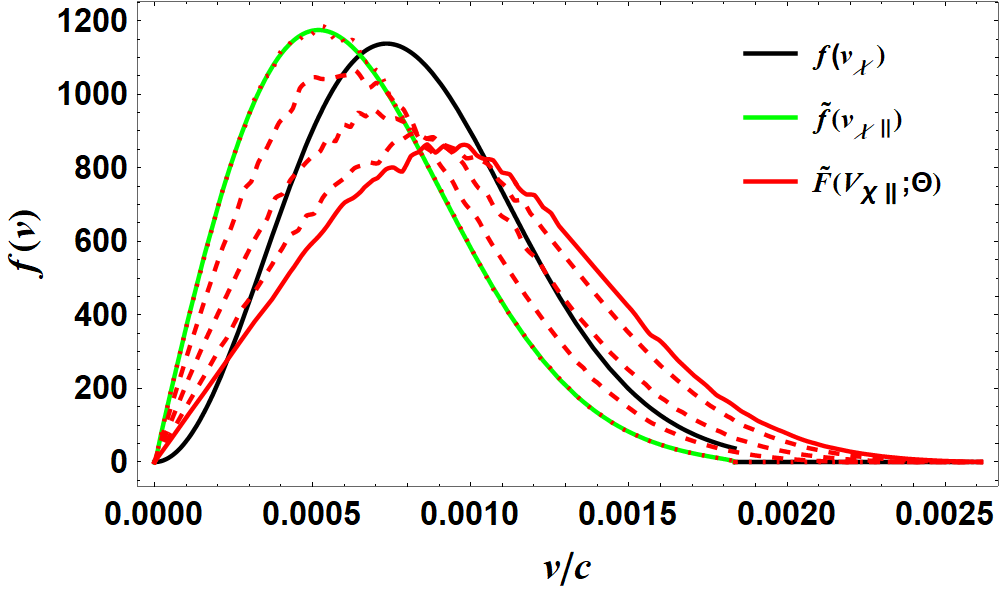}
    \caption{Velocity distributions.
    Black: the modified Maxwell-Boltzmann distribution $f(v_\chi)$, Green: the plane-projected modified Maxwell-Boltzmann distribution $\tilde{f}(v_{\chi\parallel})$, Red: the plane-projected velocity profile $\tilde{F}(V_{\chi \parallel};\Theta)$ with $\Theta = \dfrac{i}{10} \dfrac{\pi}{2}, (i=10, 7, 5, 3, 0~[{\rm solid~to~dotted}])$.}
    \label{fig:Velocity}
\end{figure}

For comparison purposes, in FIG.~\ref{fig:Velocity}, we display the modified Maxwell-Boltzmann distribution in Eq.~\eqref{eq:ModMB} (black line), the plane-projected modified Maxwell-Boltzmann distribution in Eq.~\eqref{eq:ppMBMod} (green line), and the plane-projected velocity profile $\tilde{F}$ in Eq.~\eqref{eq:AngModMB} with several reference $\Theta$ values (red lines). 
The solid red line corresponds to $\Theta = \dfrac{\pi}{2}$, where the net dark-matter flux lies within the two-dimensional detector plane, while the dotted red line represents $\Theta = 0$, where the net dark-matter flux is aligned with the detector plane-normal direction.
As expected, if the dark-matte wind blows in the direction parallel (perpendicular) to the detector plane, the velocity distribution becomes harder (softer). 

\medskip

\noindent{\bf Application to the GJJ detector.}
We now apply the formalism developed in the previous section to the GJJ-based superlight dark-matter detector proposed in Ref.~\cite{Kim:2020coy} and demonstrate that $n_{\rm eve}(\Theta)$ enables the determination of $m_\chi$ even without access to differential recoil energy spectra.

Following the calculation details in Ref.~\cite{Kim:2020coy} and assuming dark matter-electron scattering, we find that $\overline{N}_T$ is modified to 
\begin{equation}
    \overline{N}_e = \frac{n_e^{\rm 2D}}{\rho_{\rm gr}^{\rm 2D}}\,,
\end{equation}
where $\rho_{\rm gr}^{\rm 2D}$ and $n_e^{\rm 2D}$ are the graphene areal density, for which we take $\rho_{\rm gr}^{\rm 2D}=7.62\times 10^{-8}$~g/cm$^2$, and the free-electron number density on a graphene sheet, respectively. 
For consistency, we have replaced the subscript $T$ with $e$. 
The velocity-averaged event rate per unit time $\langle n^{\rm 2D}_e \sigma_{\chi e} V_{{\rm rel}\parallel} \rangle$ is
\begin{equation}
    \langle n^{\rm 2D}_e \sigma_{\chi e} V_{{\rm rel}\parallel} \rangle =\int  \frac{d^3p_{\chi,f}}{(2\pi)^3}  
    \frac{\overline{|\mathcal{M}|^2}}{16m_e^2m_\chi^2}S_{\rm ims}^2(E_r,q)S_{\rm gr}(E_r,q)\,, \label{eq:velavesig}
\end{equation}
where $p_{\chi,f}$ is the momentum of final-state dark matter, and all total energy quantities are considered in the non-relativistic limit.
Here $\overline{|\mathcal{M}|^2}$ represents the spin-averaged matrix element squared for the scattering process between dark matter and free electrons.
The remaining two factors, $S_{\rm ims}^2$ and $S_{\rm gr}$, describe the structure functions parameterizing the in-medium screening and Pauli-blocking effects, respectively. 

As per the calculational approach in Ref.~\cite{Griffin:2021znd}, the in-medium effect is given by $S_{\rm ims}(E_r,q)=(\hat{\bf q}\cdot {\bf \epsilon}\cdot \hat{\bf q})^{-1}$, where {\bf $\epsilon$} is the dielectric tensor of graphene. 
For reference, we briefly review the graphene dielectric function described in Ref.~\cite{PhysRevB.75.205418} in the Appendix. 
$S_{\rm gr}(E_r,q)$ is a function of electron-recoil kinetic energy $E_r$ and the magnitude of momentum transfer along the graphene surface, given by $q=|\Vec{p}_{\chi\parallel,i}-\Vec{p}_{\chi\parallel,f}|$, where the subscript $i$ denotes the initial state. 
We can then convert $d^3p_{\chi,f}$ to $dE_r$ and $dq$:
\begin{equation}
    \frac{d^3p_{\chi,f}}{(2\pi)^3}\, \longrightarrow\, \frac{dE_r dq}{(2\pi)^2} \frac{ 2q(E_{\chi,i}-E_r)}{\tilde{\lambda}(q^2,p_{\chi,i}^2,p_{\chi,f}^2)}\,,
\end{equation}
where $\tilde{\lambda}(x,y,z)=\sqrt{2(xy+yz+zx)-x^2-y^2-z^2}$ and we integrate out $p_{\chi,f}^z$ by extracting the delta-function factor in $S_{\rm gr}=(2\pi)\delta(p_{\chi,i}^z-p_{\chi,f}^z)\cdot S$ (see the Appendix).
The closed-form expression for $S$ was derived in Ref.~\cite{Hochberg:2015fth} based on Ref.~\cite{Reddy:1997yr}:
\begin{equation}
    S(E_r,q)=\frac{m_e^2T}{\pi q}\left[ \frac{E_r/T}{1-\exp(-E_r/T)}\left(1+\frac{\xi}{E_r/T} \right) \right], \label{eq:Sexp}
\end{equation}
where $T$ is the system temperature surrounding the detector which we take to be 10~mK for our calculation.
The quantity $\xi$ is given by
\begin{equation}
    \xi =\log\left[ \frac{1+e^{(\varepsilon-\mu)/T}}{1+e^{(\varepsilon+E_r-\mu)/T}} \right],
\end{equation}
where $\varepsilon$ is defined as
\begin{equation}
    \varepsilon=\frac{1}{4}\frac{\lbrace E_r-q^2/(2m_e)\rbrace^2}{q^2/(2m_e)}\,,
\end{equation}
and where the chemical potential $\mu$ is identified as the Fermi energy $E_F$ at zero temperature.
The linear energy-momentum dispersion for graphene is given by $E_F=v_F\sqrt{\pi n_c}$, where the Fermi velocity of graphene is $v_F=1.15\times 10^8$~cm$\cdot$s$^{-1}$~\cite{PhysRevLett.108.116404} and carrier density is set to be $n_c=10^{12}$~cm$^{-2}$~\cite{GJJ}.  

Assuming that a dark-matter particle scatters off an electron via a $t$-channel exchange of a mediator $\phi$, the scattering cross section $\sigma_{\chi e}$ in the non-relativistic limit is given by
\begin{equation}
    \sigma_{\chi e} \approx \frac{g_\chi^2g_e^2}{\pi}\frac{\mu_{\chi e}^2}{(m_\phi^2+q^2)^2}\,,\label{eq:nonrelscat}
\end{equation}
where $g_{\chi(e)}$ parameterizes the coupling strength of $\phi$ to $\chi$ (electron) and $\mu_{\chi e}$ denotes the reduced mass of the $\chi$-electron system. 
On the other hand, the matrix element in Eq.~\eqref{eq:velavesig} is related to the dark matter-electron scattering cross section by
\begin{equation}
    \sigma_{\chi e}=\frac{1}{16\pi} \frac{\overline{|\mathcal{M}|^2}}{m_\chi^2 m_e^2} \mu_{\chi e}^2\,.\label{eq:scat}
\end{equation}

For illustration, we consider a heavy-mediator scenario where the mediator mass $m_\phi$ is significantly larger than the momentum flow $q$, i.e., $m_\phi^2 \gg q^2$, while the application for light-mediator scenarios is straightforward. 
In this limit, the expression in \eqref{eq:nonrelscat} is further simplified to 
\begin{equation}
    \sigma_{\chi e}^{\rm heavy}\approx \frac{g_\chi^2g_e^2}{\pi}\frac{\mu_{\chi e}^2}{m_\phi^4}\,. \label{eq:heavy}
\end{equation}
Equating this and Eq.~\eqref{eq:scat}, we obtain the matrix element squared for heavy-mediator scenarios:
\begin{equation}
    \overline{|\mathcal{M}|^2}_{\rm heavy}=16 g_\chi^2 g_e^2\frac{m_\chi^2 m_e^2}{m_\phi^4}\,.
\end{equation}
The event rate is then determined by substituting this expression into Eq.~\eqref{eq:master} through Eq.~\eqref{eq:velavesig}.

\begin{figure}[t]
    \centering
    \includegraphics[width=0.48\textwidth]{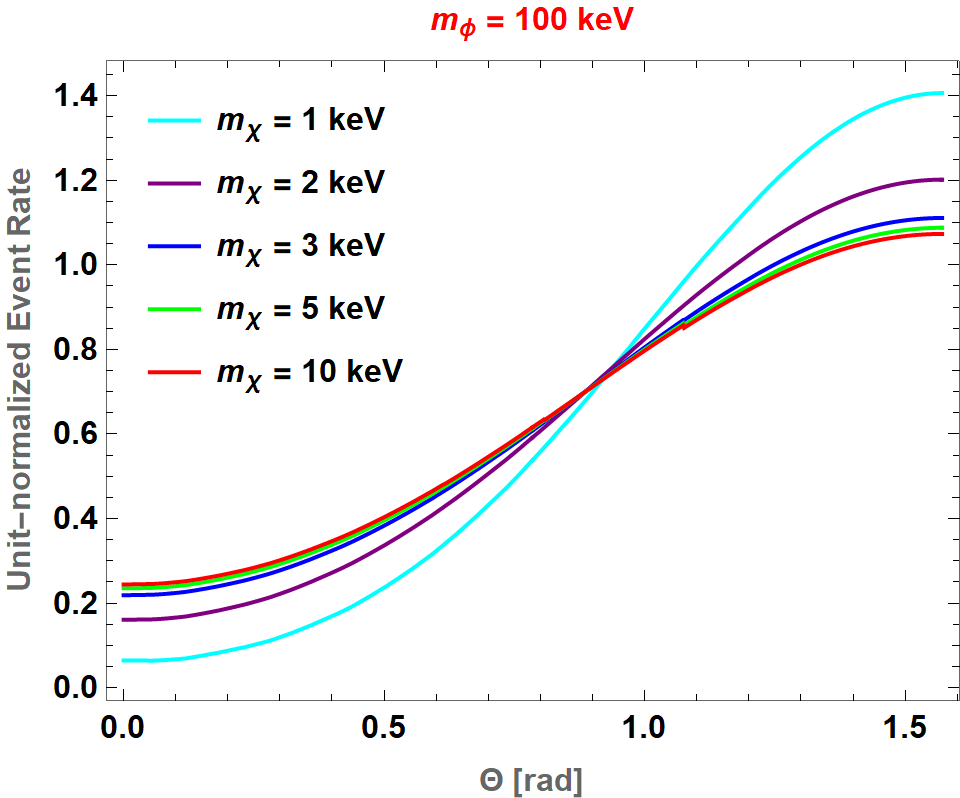}
    \caption{Unit-normalized event rates for various dark-matter mass values as a function of $\Theta$. A heavy-mediator scenario is considered with $m_\phi=100$~keV and $E_{\rm th}=1$~meV.    
    }
    \label{fig:EventRate}
\end{figure}

We present the unit-normalized event rates for various dark-matter mass values as a function of $\Theta$ in FIG.~\ref{fig:EventRate}, assuming $m_\phi=100$~keV and $E_{\rm th}=1$~meV.  
As we discussed earlier, lower-mass dark matter carries less energy, making the associated signal detection efficiency more strongly dependent on the incident angle relative to the graphene plane.
Figure~\ref{fig:EventRate} unambiguously supports this expectation, showing the strongest angular dependence for $m_\chi=1$~keV (cyan line) and the weakest for $m_\chi = 10$~keV (red line), respectively. 

In an experimental setup, one can rotate the detector plane to vary the $\Theta$ value, measure the event rates for a fixed exposure, and then compare the measured data with the predictions shown in FIG.~\ref{fig:EventRate}. 
In practice, however, modifying an established experimental configuration may introduce unwanted noise and necessitate additional calibration and stabilization procedures, thereby creating new sources of uncertainty. 
To avoid these complications, $\Theta$ can instead be inferred from the event time and the detector's geographical location when a dark-matter signal candidate is recorded---effectively realizing a virtual rotation. 
To this end, we have developed a code package {\it DarkWind} to evaluate the direction of dark-matter wind and $\Theta$ by accounting for the motions of the Sun and the Earth. 
The associated uncertainty is obtained by propagating the uncertainties in the Galactic rotation velocity and the Sun’s peculiar velocity, yielding an uncertainty in $\Theta$ at the level of $\sim 0.33^\circ$.
For detailed instructions and additional applications of the code, see the Supplemental Material. 

\medskip

\noindent {\bf Conclusions.} 
In this work, we proposed a novel method for determining the mass scale of ambient dark matter using directionality observables in (effectively) two-dimensional direct detection experiments. 
By leveraging the angular dependence of event rates induced by the directional dark-matter flux, we demonstrated that the dark-matter mass can be extracted even in the absence of differential recoil energy spectra. 
Our numerical results, based on the GJJ-based detector, validate these theoretical expectations and highlight the feasibility of this approach. 
To infer the dark-matter mass, a simple likelihood fit of the binned $\Theta$-distribution can be used, treating the overall normalization as a nuisance parameter. 
Given its broad applicability to other low-threshold (two-dimensional) directional detectors, this method opens new avenues for probing superlight dark matter and refining mass determination techniques in future experiments.

\medskip

\section*{Acknowledgments} 
We thank Hongki Min for numerous useful discussions, regarding the dielectric function of graphene. 
The work was supported by the Samsung Science and Technology Foundation (Project No. SSTF-BA2101-06).
This work was supported by the National Research Foundation of Korea (NRF) grant funded by the Ministry of Science and ICT (RS-2024-00356960) and by Global - Learning \& Academic research Institution for Master's · PhD students, and Postdocs (G-LAMP) Program of the NRF grant funded by the Ministry of Education (RS-2025-25442707). 
 
\onecolumngrid
\section*{Appendix}

\noindent {\bf Dielectric Function of Graphene.} 
We summarize the calculation done in Ref.~\cite{PhysRevB.75.205418}. 
In the random phase approximation, the dynamical dielectric function is given by
\begin{equation}
    \epsilon(E_r,q)=1+v_e(q)\Pi(E_r,q)\,,
\end{equation}
where $v_e(q)=\frac{8\pi^2\alpha_{\rm em}}{\kappa q}$ is the two-dimensional Coulomb interaction with the electromagnetic fine structure constant $\alpha_{\rm em}$ and the background lattice dielectric constant $\kappa$ set to be 4 (using SiO$_2$ as the substrate material) and $\Pi$ is the two-dimensional polarizability given by  
\begin{equation}
    \Pi(E_r,q)=\sqrt{\frac{n_c}{\pi}}\frac{1}{v_F}\tilde{\Pi}(E_r,q)\,.
\end{equation}
The dimensionless polarizability $\tilde{\Pi}$ is decomposed into
\begin{equation}
    \tilde{\Pi}(\nu,x)=\tilde{\Pi}^+(\nu,x)+\tilde{\Pi}^-(\nu,x)
\end{equation}
as a function of dimensionless quantities $\nu\equiv E_r/E_F$ and $x=q/k_F$ with $k_F=E_F/v_F$ being the Fermi momentum of graphene. 
$\tilde{\Pi}^+$ is further decomposed into
\begin{equation}
    \tilde{\Pi}^+(\nu,x)=\tilde{\Pi}_1^+(\nu,x)\theta(\nu-x)+\tilde{\Pi}_2^+(\nu,x)\theta(x-\nu)\,,
\end{equation}
where $\theta(\cdots)$ represents the usual Heaviside step function. 
Both $\tilde{\Pi}_1^+$ and $\tilde{\Pi}_2^+$ have real and imaginary parts, i.e., $\tilde{\Pi}_{1,2}^+=\Re \tilde{\Pi}_{1,2}^+ +i \Im \tilde{\Pi}_{1,2}^+$, which are expressed as
\begin{eqnarray}
    \Re \tilde{\Pi}_1^+(\nu,x) &=&1-\frac{1}{8\sqrt{\nu^2-x^2}} \left\{ f_1(\nu,x)\theta(|2+\nu|-x) +{\rm sgn}(\nu-2+x)f_1(-\nu,x)\theta(|2-\nu|-x) \right. \nonumber \\
    &+& \left. f_2(\nu,x)\left[\theta(x+2-\nu)+\theta(2-x-\nu) \right]\right\}, \\
    \Im \tilde{\Pi}_1^+(\nu,x) &=& \frac{-1}{8\sqrt{\nu^2-x^2}}\left\{ f_3(-\nu,x)\theta(x-|\nu-2|)+\frac{\pi x^2}{2}\left[\theta(x+2-\nu) +\theta(2-x-\nu) \right] \right\}, \\
    \Re \tilde{\Pi}_2^+(\nu,x) &=& 1-\frac{1}{8\sqrt{x^2-\nu^2}} \left\{f_3(\nu,x)\theta(x-|\nu+2|)+f_3(-\nu,x)\theta(x-|\nu-2|)+\frac{\pi x^2}{2}\left[\theta(|\nu+2|-x) +\theta(|\nu-2|-x) \right] \right\}, \nonumber \\
    && \\
    \Im \tilde{\Pi}_2^+(\nu,x) &=& \frac{\theta(\nu-x+2)}{8\sqrt{x^2-\nu^2}} \left\{f_4(\nu,x)-f_4(-\nu,x)\theta(2-x-\nu) \right\},
\end{eqnarray}
where ${\rm sgn}(\cdots)$ is the usual sign function and $f_{1,2,3,4}$ are defined as
\begin{eqnarray}
    f_1(\nu,x)&=&(2+\nu)\sqrt{(2+\nu)^2-x^2}-x^2\log \frac{\sqrt{(2+\nu)^2-x^2}+(2+\nu)}{|\sqrt{\nu^2-x^2}+\nu|}\,, \\
    f_2(\nu,x)&=& x^2 \log \frac{\nu-\sqrt{\nu^2-x^2}}{x}\,, \\
    f_3(\nu,x)&=& (2+\nu)\sqrt{x^2-(2+\nu)^2}+x^2\sin^{-1}\frac{2+\nu}{x}\,, \\
    f_4(\nu,x)&=& (2+\nu)\sqrt{(2+\nu)^2-x^2}-x^2\log \frac{\sqrt{(2+\nu)^2-x^2}+(2+\nu)}{x}\,.
\end{eqnarray}
On the other hand, $\tilde{\Pi}^-$ is simply given by
\begin{equation}
    \tilde{\Pi}^-(\nu,x)=\Re\tilde{\Pi}^-(\nu,x) +i \Im \tilde{\Pi}^-(\nu,x)= \frac{\pi x^2\theta(x-\nu)}{8\sqrt{x^2-\nu^2}}+i \frac{\pi x^2\theta(\nu-x)}{8\sqrt{\nu^2-x^2}}\,.
\end{equation}

\medskip

\noindent {\bf Pauli Blocking.} 
The structure function $S_{\rm gr}$ for the graphene system is expressed as
\begin{equation}
    S_{\rm gr}(E_r,q)=2 \int \frac{d^3p_{e,i}}{(2\pi)^3}\int \frac{d^3p_{e,f}}{(2\pi)^3}(2\pi)\delta(p_{e,i}^z-p_{e,f}^z) \cdot (2\pi)^4\delta^{(4)}(p_{\chi,i}+p_{e,i}-p_{\chi,f}-p_{e,f})\cdot f_{e,i}( 1-f_{e,f})\,,
\end{equation}
where $p_{e,i(f)}$ denotes the four-momentum of the initial-state (final-state) electron, and $p_{e,i(f)}^z$ represents its $z$-component spatial momentum.  
The first delta function reflects the assumption that free electrons are confined to the graphene surface, meaning that they experience negligible momentum change along the surface-normal direction, provided that the electron recoil kinetic energy remains lower than the graphene work function.
The Fermi-Dirac distribution functions for the initial-state (final-state) electrons $f_{e,i(f)}$ are given by
\begin{equation}
    f_{e,i(f)}=\left\{1+\exp\left(\frac{E_{e,i(f)}-\mu}{T} \right) \right\}^{-1}\,,
\end{equation}
where $E_{e,i(f)}$ represents the energy of the initial-state (final-state) electron and where $\mu$ and $T$ denote the chemical potential and the system temperature, respectively.
By integrating over $d^3p_{e,f}$ along with the spatial components of the four-dimensional delta function, we obtain
\begin{eqnarray}
    S_{\rm gr}(E_r,q)&=&(2\pi)\delta(p_{\chi,i}^z-p_{\chi,f}^z)\cdot \frac{1}{2\pi^2}\int d^3p_{e,i}  \delta(E_r+E_{\chi,i}-E_{\chi,f})f_{e,i}(1-f_{e,f}) \nonumber \\ 
    &\equiv& (2\pi)\delta(p_{\chi,i}^z-p_{\chi,f}^z)\cdot S(E_r,q)\,,
\end{eqnarray}
where we factor out the delta function of $p_{\chi,f}^z$, which is utilized in the $d^3p_{\chi,f}$ integral, and define $S(E_r,q)$ separately. 
As shown in the main text, a closed-form expression for $S(E_r,q)$ is available in the non-relativistic limit~\cite{Reddy:1997yr}:
\begin{equation}
    S(E_r,q)=\frac{m_e^2T}{\pi q}\left[ \frac{E_r/T}{1-\exp(-E_r/T)}\left(1+\frac{\xi}{E_r/T} \right) \right],
\end{equation}
where $\xi$ is given by
\begin{equation}
    \xi =\log\left[ \frac{1+e^{(\varepsilon-\mu)/T}}{1+e^{(\varepsilon+E_r-\mu)/T}} \right],
\end{equation}
where $\varepsilon$ is defined as
\begin{equation}
    \varepsilon=\frac{1}{4}\frac{\lbrace E_r-q^2/(2m_e)\rbrace^2}{q^2/(2m_e)}\,.
\end{equation}

\newpage


\noindent {\bf Direction of Dark-Matter Wind.} We begin by defining the quantities used in this section and listing their numerical input values in Table~\ref{tab:symbol}.

\begin{table}[h]
\small
\setlength{\tabcolsep}{4pt}
\renewcommand{\arraystretch}{1.0}
\begin{tabular}{c | c | c}
\hline \hline
Symbol & Definition & Value \\
\hline

\begin{minipage}[t]{0.1\columnwidth}\vspace{0pt} 
\begin{center}
    $v_c$
\end{center}
\end{minipage}
& \begin{minipage}[t]{0.42\columnwidth}\vspace{0pt}
Galactic rotation speed (local circular speed)
\end{minipage}
& \begin{minipage}[t]{0.43\columnwidth}\vspace{0pt}
$v_c=218~{\rm km/s} ~ \pm ~ 6~{\rm km/s}$ \cite{Bovy:2012tw}
\end{minipage}
\\

\begin{minipage}[t]{0.1\columnwidth}\vspace{0pt} \centering $U,V,W$
\end{minipage}
& \begin{minipage}[t]{0.42\columnwidth}\vspace{0pt}
Components of the Sun's peculiar velocity
\end{minipage}
& \begin{minipage}[t]{0.43\columnwidth}\vspace{0pt}
$\displaystyle (U,V,W)_\odot = \Big(11.1^{+0.69}_{-0.75},\,12.24^{+0.47}_{-0.47},\,7.25^{+0.37}_{-0.36}\Big)\ {\rm km/s}$ \cite{Schoenrich:2010}
\end{minipage}
\\

\begin{minipage}[t]{0.1\columnwidth}\vspace{0pt} \centering $v_\oplus$
\end{minipage}
& \begin{minipage}[t]{0.42\columnwidth}\vspace{0pt}
Magnitude of the Earth's orbital speed
\end{minipage}
& \begin{minipage}[t]{0.43\columnwidth}\vspace{0pt}
$v_\oplus=29.79~{\rm km/s}$ \cite{AstronomicalAlmanac2015}
\end{minipage}
\\

\begin{minipage}[t]{0.1\columnwidth}\vspace{0pt} \centering $e$
\end{minipage}
& \begin{minipage}[t]{0.42\columnwidth}\vspace{0pt}
Orbital eccentricity
\end{minipage}
& \begin{minipage}[t]{0.43\columnwidth}\vspace{0pt}
$e=0.0167023$ \cite{AstronomicalAlmanac2015}
\end{minipage}
\\

\begin{minipage}[t]{0.1\columnwidth}\vspace{0pt} \centering $L$
\end{minipage}
& \begin{minipage}[t]{0.42\columnwidth}\vspace{0pt}
Mean longitude of the Sun
\end{minipage}
& \begin{minipage}[t]{0.43\columnwidth}\vspace{0pt}
$L = 279^\circ.344 + 0.9856474\,d$ \cite{AstronomicalAlmanac2015}
\end{minipage}
\\

\begin{minipage}[t]{0.1\columnwidth}\vspace{0pt} \centering $g$
\end{minipage}
& \begin{minipage}[t]{0.42\columnwidth}\vspace{0pt}
Mean anomaly of the Earth's orbit
\end{minipage}
& \begin{minipage}[t]{0.43\columnwidth}\vspace{0pt}
$g = 356^\circ.154 + 0.9856003\,d$ \cite{AstronomicalAlmanac2015}
\end{minipage}
\\

\begin{minipage}[t]{0.1\columnwidth}\vspace{0pt} \centering $d$
\end{minipage}
& \begin{minipage}[t]{0.42\columnwidth}\vspace{0pt}
Fractional day number used to compute $L$ and $g$ \\
(from Dec.\ 31, 2014 at 0:00 UT)
\end{minipage}
& \begin{minipage}[t]{0.43\columnwidth}\vspace{0pt}
$\displaystyle
\begin{aligned}
d&=\Big\lfloor 365.25\,\tilde{Y}\Big\rfloor + \Big\lfloor 30.61(\tilde{M}+1)\Big\rfloor + D \\
&\quad + \frac{UT}{24} - 736041 ,
\end{aligned}
$ \cite{Meeus:1998}\\
For $M=1,2$: $\tilde{Y}=Y-1$, $\tilde{M}=M+12$; otherwise $\tilde{Y}=Y$, $\tilde{M}=M$.\\
($Y$: year, $M$: month, $D$: day, $UT$: universal time in hours.)
\end{minipage}
\\

\begin{minipage}[t]{0.1\columnwidth}\vspace{0pt} \centering $T$
\end{minipage}
& \begin{minipage}[t]{0.42\columnwidth}\vspace{0pt}
Epoch of date (precession parameter)
\end{minipage}
& \begin{minipage}[t]{0.43\columnwidth}\vspace{0pt}
$T=d/36525$
\end{minipage}
\\

\begin{minipage}[t]{0.1\columnwidth}\vspace{0pt} \centering $(\beta_i,\lambda_i)$
\end{minipage}
& \begin{minipage}[t]{0.42\columnwidth}\vspace{0pt}
Ecliptic latitude/longitude of the galactic axes \\
($i=x,y,z$)
\end{minipage}
& \begin{minipage}[t]{0.43\columnwidth}\vspace{0pt}
$\displaystyle
\begin{aligned}
(\beta_x,\lambda_x) &= (5^\circ.538,\,267^\circ.050) + (0^\circ.013,\,1^\circ.397)\,T\,,\\
(\beta_y,\lambda_y) &= (-59^\circ.574,\,347^\circ.546) + (0^\circ.002,\,1^\circ.375)\,T\,,\\
(\beta_z,\lambda_z) &= (29^\circ.811,\,180^\circ.234) + (0^\circ.001,\,1^\circ.404)\,T\,.
\end{aligned}
$ \cite{Mayet:2016sva, McCabe:2014gka}
\end{minipage}
\\

\begin{minipage}[t]{0.1\columnwidth}\vspace{0pt} \centering $v^{\rm eq}_{e,{\rm rot}}$
\end{minipage}
& \begin{minipage}[t]{0.42\columnwidth}\vspace{0pt}
Earth's rotational speed at the equator
\end{minipage}
& \begin{minipage}[t]{0.43\columnwidth}\vspace{0pt}
$v^{\rm eq}_{e,{\rm rot}}=0.465102~{\rm km/s}$ \cite{Mayet:2016sva}
\end{minipage}
\\

\begin{minipage}[t]{0.1\columnwidth}\vspace{0pt} \centering $\lambda_{\rm lab}$ 
\end{minipage}
& \begin{minipage}[t]{0.42\columnwidth}\vspace{0pt}
Laboratory latitude
\end{minipage}
& \begin{minipage}[t]{0.43\columnwidth}\vspace{0pt}
(site-dependent)
\end{minipage}
\\

\begin{minipage}[t]{0.1\columnwidth}\vspace{0pt} \centering $t^\circ_{\rm lab}$ 
\end{minipage}
& \begin{minipage}[t]{0.42\columnwidth}\vspace{0pt}
Local sidereal time
\end{minipage}
& \begin{minipage}[t]{0.43\columnwidth}\vspace{0pt}
(time-dependent)
\end{minipage}
\\

\begin{minipage}[t]{0.1\columnwidth}\vspace{0pt} \centering $\mathbf A_G$
\end{minipage}
& \begin{minipage}[t]{0.42\columnwidth}\vspace{0pt}
Galactic-to-equatorial rotation matrix
\end{minipage}
& \begin{minipage}[t]{0.43\columnwidth}\vspace{0pt}
see Eq.~(\ref{eq:AG}) \cite{Hipparcos:1997}  
\end{minipage}
\\ 

\hline \hline
\end{tabular}
\caption{Definitions of symbols and their numerical input values.} \label{tab:symbol}
\end{table}

In the main text, for simplicity, we define $\Theta$ as the angle between the Cygnus direction and the detection-plane-normal direction. 
In an actual experiment, however, $\Theta$ is given by the angle between the dark-matter wind direction and the detection-plane-normal direction.
In general, if we denote the unit normal vector of the detector plane in the laboratory frame by $\hat{\mathbf n}_{\rm det}$ and the unit vector of the dark-matter wind direction in the laboratory frame by $\hat{\mathbf v}_{\rm wind}(t) = -\hat{\mathbf v}_{\odot}$, then
\begin{equation}
{\Theta(t)=\arccos\big(\hat{\mathbf n}_{\rm det}\cdot \hat{\mathbf v}_{\rm wind}(t))}
\end{equation}
defines $\Theta(t)$.
In particular, when the detector plane is parallel to the horizontal plane and the wind direction has altitude $\alpha$,
\begin{equation}
\Theta = \frac{\pi}{2}-\alpha\,.
\end{equation}

The dark-matter wind arises from the motion of the Sun and the Earth within the Milky Way. 
The coordinate conventions and the calculation procedure used below follow Appendix~A of Ref.~\cite{Mayet:2016sva}, except for some updated input values.
The galactic frame $(\hat{\mathbf x}_g,\hat{\mathbf y}_g,\hat{\mathbf z}_g)$ used to combine the velocity vectors of the Sun and the Earth takes the Sun as the origin, and its axes point toward the Galactic center, the direction of Galactic rotation, and the north Galactic pole, respectively.
The equatorial frame $(\hat{\mathbf x}_e,\hat{\mathbf y}_e,\hat{\mathbf z}_e)$ takes the Earth as the origin, and its axes point toward the vernal equinox, the direction of right ascension $90^\circ$ along the celestial equator, and the north celestial pole, respectively.
Finally, the laboratory frame $(\hat{\mathbf N},\hat{\mathbf W},\hat{\mathbf Z})$ takes the detector as the origin, with the axes pointing toward North, West, and the local vertical (upward) direction, respectively.

We now calculate the velocity vectors of the Sun and the Earth. 
The Sun's velocity can be described as the sum of the Galactic rotation velocity and the Sun's peculiar velocity relative to it; the Galactic rotation velocity is
\begin{equation}
\mathbf v_c = v_c\,\hat{\mathbf y}_g\,,
\end{equation}
and the Sun's peculiar velocity is given by
\begin{equation}
\mathbf v_s = U\hat{\mathbf x}_g + V\hat{\mathbf y}_g + W\hat{\mathbf z}_g\,.
\end{equation}
The Earth's motion can be decomposed into revolution and rotation. 
The Earth's revolution in the galactic frame is~\cite{McCabe:2014gka}
\begin{eqnarray}
\mathbf v_{e,{\rm rev}}&=&v_\oplus\left(
\cos\beta_x[\sin(L-\lambda_x)+e\sin(L+g-\lambda_x)]\hat{\mathbf x}_g
+\cos\beta_y[\sin(L-\lambda_y)-e\sin(L+g-\lambda_y)]\hat{\mathbf y}_g \right. \nonumber \\
&&+\left.\cos\beta_z[\sin(L-\lambda_z)-e\sin(L+g-\lambda_z)]\hat{\mathbf z}_g
\right)\,,
\end{eqnarray}
while the Earth's rotation in the galactic frame is~\cite{Bozorgnia:2011vc}
\begin{eqnarray}
\mathbf v_{e,{\rm rot}}
=-v^{\rm eq}_{e,{\rm rot}}\cos\lambda_{\rm lab}\Big[
(a_x\sin t^\circ_{\rm lab}-a_y\cos t^\circ_{\rm lab})\hat{\mathbf x}_g
+(b_x\sin t^\circ_{\rm lab}-b_y\cos t^\circ_{\rm lab})\hat{\mathbf y}_g
+(c_x\sin t^\circ_{\rm lab}-c_y\cos t^\circ_{\rm lab})\hat{\mathbf z}_g
\Big]\,.
\end{eqnarray}
The coefficients $(a_x, \ldots, c_z)$ are defined as the elements of the galactic-to-equatorial rotation matrix $\mathbf A_G$~\cite{Hipparcos:1997}:
\begin{equation}
\begin{pmatrix}
\hat{\mathbf x}_e\\
\hat{\mathbf y}_e\\
\hat{\mathbf z}_e
\end{pmatrix}
=
\mathbf A_G
\begin{pmatrix}
\hat{\mathbf x}_g\\
\hat{\mathbf y}_g\\
\hat{\mathbf z}_g
\end{pmatrix}
\hbox{ with }
\mathbf A_G=
\begin{pmatrix}
a_x & b_x & c_x\\
a_y & b_y & c_y\\
a_z & b_z & c_z
\end{pmatrix}
=
\begin{pmatrix}
-0.0548755604 & +0.4941094279 & -0.8676661490\\
-0.8734370902 & -0.4448296300 & -0.1980763734\\
-0.4838350155 & +0.7469822445 & +0.4559837762
\end{pmatrix}.
\label{eq:AG}
\end{equation}
Summing all velocity components, we find that the velocity vector of the detector of interest with respect to the galactic frame, ${\bf v_{\rm det,gal}}$, is given by
\begin{align}
\mathbf v_{\rm det,gal} &= {\bf v}_c + {\bf v}_s + {\bf v}_{\rm e, rev}+{\bf v}_{\rm e,rot} \nonumber \\
&=
\Big(U+v_\oplus\cos\beta_x[\sin(L-\lambda_x)+e\sin(L+g-\lambda_x)]
-v^{\rm eq}_{e,{\rm rot}}\cos\lambda_{\rm lab}(a_x\sin t^\circ_{\rm lab}-a_y\cos t^\circ_{\rm lab})\Big)\hat{\mathbf x}_g
\nonumber\\
&\quad+\Big(v_c+V+v_\oplus\cos\beta_y[\sin(L-\lambda_y)-e\sin(L+g-\lambda_y)]
-v^{\rm eq}_{e,{\rm rot}}\cos\lambda_{\rm lab}(b_x\sin t^\circ_{\rm lab}-b_y\cos t^\circ_{\rm lab})\Big)\hat{\mathbf y}_g
\nonumber\\
&\quad+\Big(W+v_\oplus\cos\beta_z[\sin(L-\lambda_z)-e\sin(L+g-\lambda_z)]
-v^{\rm eq}_{e,{\rm rot}}\cos\lambda_{\rm lab}(c_x\sin t^\circ_{\rm lab}-c_y\cos t^\circ_{\rm lab})\Big)\hat{\mathbf z}_g\,.
\end{align}

We now transform $\mathbf v_{\rm det,gal}$ to the laboratory frame in which the experiment is performed. 
The laboratory-to-equatorial transformation is~\cite{Bozorgnia:2011vc}
\begin{align}
\hat{\mathbf N}&= -\sin(\lambda_{\rm lab})\Big[\cos(t^\circ_{\rm lab})\hat{\mathbf x}_e+\sin(t^\circ_{\rm lab})\hat{\mathbf y}_e\Big]+\cos(\lambda_{\rm lab})\hat{\mathbf z}_e\,,\\
\hat{\mathbf W}&= \sin(t^\circ_{\rm lab})\hat{\mathbf x}_e-\cos(t^\circ_{\rm lab})\hat{\mathbf y}_e\,,\\
\hat{\mathbf Z}&= \cos(\lambda_{\rm lab})\Big[\cos(t^\circ_{\rm lab})\hat{\mathbf x}_e+\sin(t^\circ_{\rm lab})\hat{\mathbf y}_e\Big]+\sin(\lambda_{\rm lab})\hat{\mathbf z}_e\,,
\end{align}
and the equatorial-to-galactic transformation (the inverse of $\mathbf A_G$) is~\cite{Hipparcos:1997}
\begin{align}
\hat{\mathbf x}_g&=a_x\,\hat{\mathbf x}_e+a_y\,\hat{\mathbf y}_e+a_z\,\hat{\mathbf z}_e\,,\\
\hat{\mathbf y}_g&=b_x\,\hat{\mathbf x}_e+b_y\,\hat{\mathbf y}_e+b_z\,\hat{\mathbf z}_e\,,\\
\hat{\mathbf z}_g&=c_x\,\hat{\mathbf x}_e+c_y\,\hat{\mathbf y}_e+c_z\,\hat{\mathbf z}_e\,.
\end{align}
Therefore, the galactic-to-laboratory transformation can be written as
\begin{equation}
\begin{pmatrix} v_{\hat N}\\ v_{\hat W}\\ v_{\hat Z}\end{pmatrix}
=\mathbf R
\begin{pmatrix} v_{x_g}\\ v_{y_g}\\ v_{z_g}\end{pmatrix},
\end{equation}
where $(v_{x_g},v_{y_g},v_{z_g})$ are the components of $\mathbf v_{\rm det}$ along $(\hat{\mathbf x}_g,\hat{\mathbf y}_g,\hat{\mathbf z}_g)$ and $(v_{\hat N},v_{\hat W},v_{\hat Z})$ are those along $(\hat{\mathbf N},\hat{\mathbf W},\hat{\mathbf Z})$.
The transformation matrix is expressed as
\begin{equation}
\mathbf R=
\begin{pmatrix}
\begin{aligned}[t]
&-\sin\lambda_{\rm lab}\big(a_x\cos t^\circ_{\rm lab}+a_y\sin t^\circ_{\rm lab}\big)\\
&\quad +a_z\cos\lambda_{\rm lab}
\end{aligned}
&
\begin{aligned}[t]
&-\sin\lambda_{\rm lab}\big(b_x\cos t^\circ_{\rm lab}+b_y\sin t^\circ_{\rm lab}\big)\\
&\quad +b_z\cos\lambda_{\rm lab}
\end{aligned}
&
\begin{aligned}[t]
&-\sin\lambda_{\rm lab}\big(c_x\cos t^\circ_{\rm lab}+c_y\sin t^\circ_{\rm lab}\big)\\
&\quad +c_z\cos\lambda_{\rm lab}
\end{aligned}
\\[20pt]
a_x\sin t^\circ_{\rm lab}-a_y\cos t^\circ_{\rm lab}
&
b_x\sin t^\circ_{\rm lab}-b_y\cos t^\circ_{\rm lab}
&
c_x\sin t^\circ_{\rm lab}-c_y\cos t^\circ_{\rm lab}
\\[10pt]
\begin{aligned}[t]
&\cos\lambda_{\rm lab}\big(a_x\cos t^\circ_{\rm lab}+a_y\sin t^\circ_{\rm lab}\big)\\
&\quad +a_z\sin\lambda_{\rm lab}
\end{aligned}
&
\begin{aligned}[t]
&\cos\lambda_{\rm lab}\big(b_x\cos t^\circ_{\rm lab}+b_y\sin t^\circ_{\rm lab}\big)\\
&\quad +b_z\sin\lambda_{\rm lab}
\end{aligned}
&
\begin{aligned}[t]
&\cos\lambda_{\rm lab}\big(c_x\cos t^\circ_{\rm lab}+c_y\sin t^\circ_{\rm lab}\big)\\
&\quad +c_z\sin\lambda_{\rm lab}
\end{aligned}
\end{pmatrix}.
\end{equation}

Using this transformation, we obtain $\mathbf v_{\rm det, lab}(t)=v_{\hat N}(t)\hat{\mathbf N}+v_{\hat W}(t)\hat{\mathbf W}+v_{\hat Z}(t)\hat{\mathbf Z}$.
The dark-matter wind velocity is defined as the opposite of $\mathbf v_{\rm det,lab}$:
\begin{equation}
\mathbf v_{\rm wind}(t)=-\mathbf v_{\rm det,lab}(t)\,.
\end{equation}
Writing the wind-direction unit vector $\hat{\mathbf v}_{\rm wind}(t) = \mathbf v_{\rm wind}(t) / |\mathbf v_{\rm wind}(t)|$ in the laboratory-frame basis as
\begin{equation}
\hat{\mathbf v}_{\rm wind}(t)
=\hat v_{\hat N}(t)\hat{\mathbf N}
+\hat v_{\hat W}(t)\hat{\mathbf W}
+\hat v_{\hat Z}(t)\hat{\mathbf Z}\,,
\end{equation}
we introduce the east component $\hat v_{\hat E}(t) \equiv -\hat v_{\hat W}(t)$ to follow the standard azimuth convention (measured east of North).

The altitude $\alpha$ is defined by
\begin{equation}
\alpha(t)=\arcsin \big(\hat v_{\hat Z}(t)\big)\,,
\qquad \alpha\in\left[-\frac{\pi}{2},\frac{\pi}{2}\right]\,.
\end{equation}
For the azimuth $\phi$ (measured east of North, i.e. ${\rm North}=0^\circ$ and ${\rm East}=90^\circ$),
\begin{equation}
\phi(t)=
\begin{cases}
\arccos\left(\dfrac{\hat v_{\hat N}(t)}{\sqrt{\hat v_{\hat N}^2(t)+\hat v_{\hat E}^2(t)}}\right),
& \hat v_{\hat E}(t)\ge 0\,, \\[10pt]
2\pi-\arccos\left(\dfrac{\hat v_{\hat N}(t)}{\sqrt{\hat v_{\hat N}^2(t)+\hat v_{\hat E}^2(t)}}\right),
& \hat v_{\hat E}(t)< 0\,,
\end{cases}
\qquad \phi\in[0,2\pi)\,.
\end{equation}
For a detector parallel to the horizontal plane, $\Theta$ is complementary to the altitude, and thus
\begin{equation}
\Theta=\frac{\pi}{2}-{\alpha}\,.
\end{equation}

\medskip

\begin{figure}[t]
    \centering
    \includegraphics[width=0.5 \linewidth]{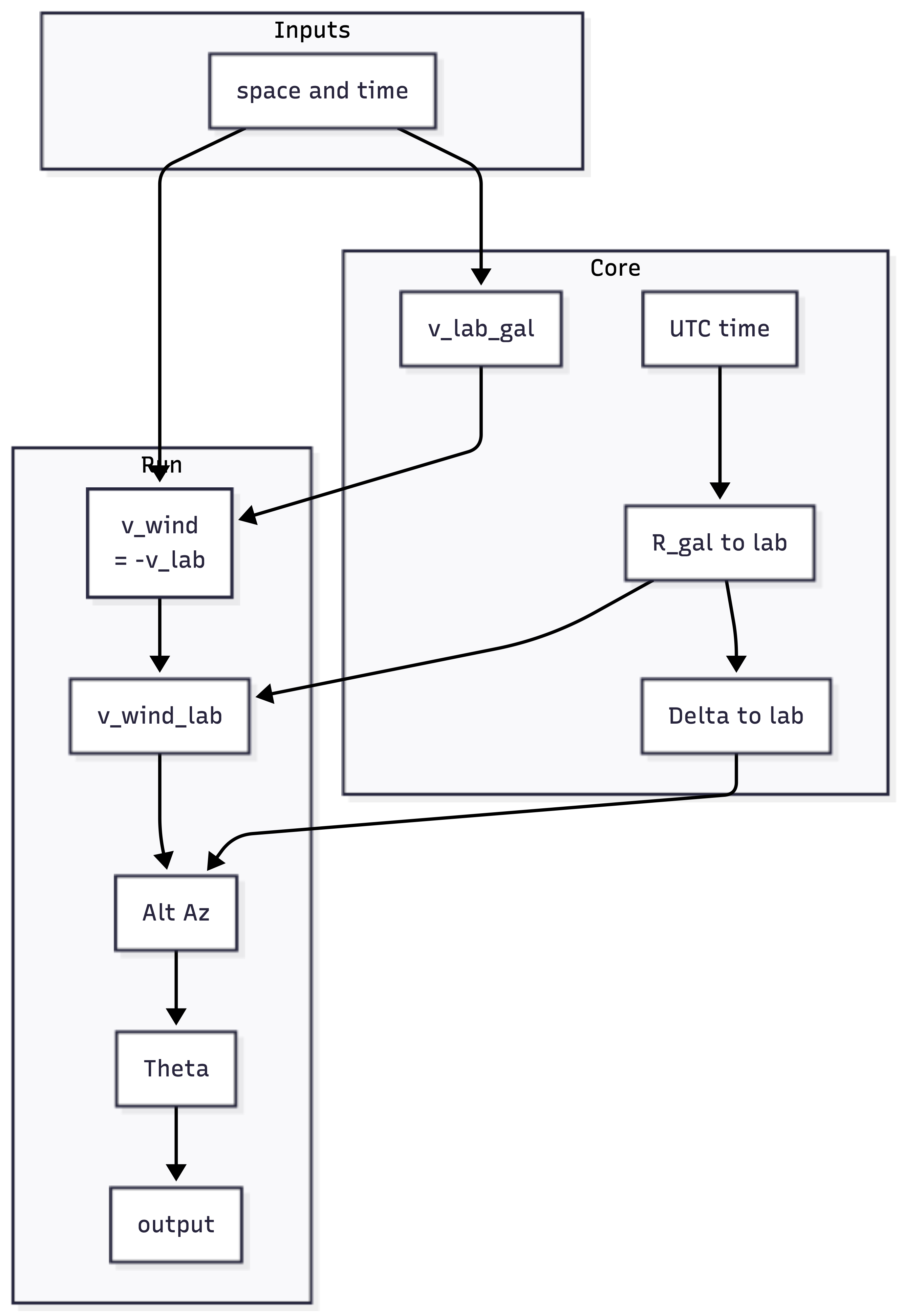} 
    \vspace{0.5cm}
    \caption{Flow chart of {\it DarkWind}.}
    \label{fig:flowchart}
\end{figure}

\noindent {\bf DarkWind: a package to search for the angle between the dark-matter wind and the detector plane.} 
We have developed a Python package that performs the above calculation and returns $\Theta$; a flowchart of the procedure is shown in FIG.~\ref{fig:flowchart}. 
Given the time, latitude, and longitude in the laboratory frame, the package provides the velocity vectors of the Sun/Earth in the galactic and laboratory frames, the dark-matter wind velocity vector in the laboratory frame, and the wind direction expressed in the horizontal coordinate system (altitude, azimuth). 
In addition, for a detector assumed to be parallel to the horizontal plane, it provides the final value of $90^\circ-{\rm altitude}$ (in degrees), which corresponds to the angle between the dark-matter wind and the detection-plane-normal vector.

The package is divided into the main calculation module \texttt{DM\_velocity\_calculator.py} and the output runner \texttt{run\_DM\_velocity\_calculator.py}.
To run the code, one can use
\begin{verbatim}
python run_DM_velocity_calculator.py --lat 00 --lon 00 --datetime 0000-00-00T00:00:00 --tz 0
\end{verbatim}
where \texttt{lat} is the detector latitude, \texttt{lon} is the longitude, \texttt{datetime} is the date and time, and \texttt{tz} is the timezone.
For example, to calculate $\Theta$ at a specific site, one can use (replace the coordinates and time with those of the desired experimental location and detection time):
\begin{verbatim}
python run_DM_velocity_calculator.py --lat 37.5666805 --lon 126.9784147 --datetime \ 
       2025-12-19T11:00:00 --tz 9
\end{verbatim}
More detailed usage can be found via executing
\begin{verbatim}
python DM_velocity_calculator.py --help
\end{verbatim}

\twocolumngrid

\bibliography{ref}

\end{document}